\documentclass[twocolumn,showpacs,preprintnumbers,amsmath,amssymb]{revtex4}
\usepackage{graphicx}% Include figure files
\usepackage{dcolumn}% Align table columns on decimal point
\usepackage{bm}% bold math

\begin{document}

%\preprint{APS/123-QED}

\title{Recurrence Tracking Microscope}% Force line breaks with \\

\author{Farhan Saif}
 \email{saif@physics.arizona.edu}
\affiliation{Department of Electronics, Quaid-i-Azam University,
Islamabad 45320, Pakistan.\\
Department of Physics, The University of Arizona, Tucson, Arizona
85721, USA.}

%\affiliation{

\date{\today}% It is always \today, today,

\begin{abstract}
In order to probe nanostructures on a surface we present a microscope based on the quantum recurrence phenomena. A cloud of atoms bounces off an atomic mirror connected to a cantilever and exhibits quantum recurrences. The times at which the
recurrences occur depend on the initial height of the bouncing
atoms above the atomic mirror, and vary following the structures
on the surface under investigation. The microscope has inherent advantages over existing techniques of scanning tunneling
microscope and atomic force microscope. Presently available
experimental technology makes it possible to develop the device in
the laboratory.\end{abstract}

\maketitle

\section{\label{sec:level1}INTRODUCTION}

In 1982, Binnig and Rohrer used quantum tunneling phenomenon as a
probe to study nano-structures on a surface~\cite{binn1}. The idea
led to the development of Scanning Tunneling microscope (STM) and
won the inventors Nobel prize in physics in 1986. Later, the
inventors of the STM designed the Atomic Force microscope
(AFM)~\cite{binn2}. In this paper we present a microscope based on quantum recurrence phenomena to probe nano-structures on a surface with
atomic size resolution, and therefore appropriately name the device as
Recurrence tracking microscope (RTM).

%However, the device has some inherent limitations, firstly, it
%requires the surface under study to be a conducting surface and,
%secondly, any surface impurity appeared as surface structure due to
%change in tunneling current. In order to avoid these deficiencies,
%
%However, based on all classical optics principles this device is
%having a limited resolution and magnification as compared with STM.

%Recurrence Tracking microscope (RTM) is based on quantum principle
%as in case of scanning tunneling microscope (STM).
%The workhorse of Recurrence Tracking Microscope (RTM) is quantum
%recurrence phenomena~\cite{saifPR}. A cloud of altra cold atoms is
%set to move on an optical surface~\cite{ovch} attached to a
%cantilever. The long time evolution of the cloud displays
%recurrences~\cite{chen1} which vary in time subject to cantilever
%position.

Recurrence Tracking Microscope is in many ways advantageous over
existing techniques of STM and AFM:
%{\it \ (i)} With the help of RTM we probe periodic surfaces and rough surfaces;
({\it i}) It probes material surfaces of all kinds ranging from
conductors to insulators; ({\it ii}) It investigates surfaces
comprising impurities without observing the impurity atoms as extra
surface structures. STM, however, has a drawback as it observes
extra surface structures for the impurity atoms~\cite{meyer};
%{\it (iii)} RSM is having higher resolution as compared with AFM and comparable with STM.
({\it iii}) In dynamical operational mode, RTM provides information
about a surface with periodic structures in the simplest
manner~\cite{saifRTM}. Due to the periodicity of the structures on
the surface, the cantilever oscillates with a finite frequency. The
oscillations of the cantilever and then of the atomic mirror, appear
as a periodically changing force to the bouncing atoms. As a result
the time of recurrence is modified and the modification factor
stores the information of the periodicity of the surface structures.

Quantum recurrences of a atomic wave packet in the
absence~\cite{Robinett} and in the presence~\cite{SaifPR} of
an oscillating surface are well understood. The phenomena
have been realized as well experimentally~\cite{kn:park}. The
increased dynamical stability in the surface traps needed to
develop RTM has extended the limits of the experiments from a few
bounces~\cite{kn:kase,Aminoff,EdHinds1} to even realize Bose-Einstein
condensation of atoms using an optical reflecting
surface~\cite{Rychtarik} and a magnetic
film~\cite{EdHinds2}.

In Sec. II, we introduce the experimental setup of the Recurrence tracking microscope (RTM). In Sec.III, and IV we explain the static and dynamical modes,
respectively. 
%periodic surfaces with resolution of the order of atomic size.
%
\section{THE EXPERIMENTAL MODEL}

\begin{figure}\includegraphics[scale=0.6]{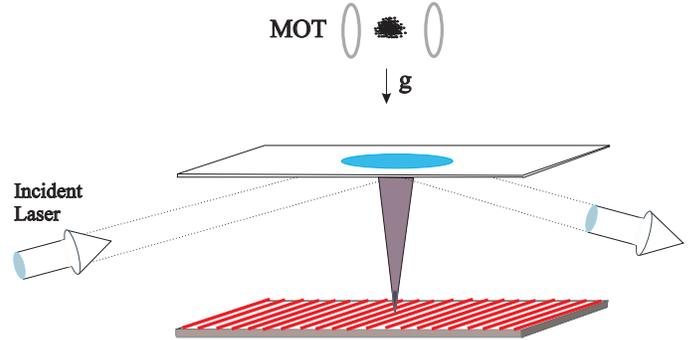}
\caption{Experimental setup of the Recurrence Tracking Microscope: A
cloud of atoms is trapped and cooled in a magneto optical trap (MOT)
down to the micro-Kelvin scale. The MOT is placed at a certain
height above the evanescent wave atomic mirror. The mirror for the
atoms results due to the total internal reflection of the incident
laser light field from the surface of the dielectric film. Thus an
evanescent wave field is produced on the surface which has
polarization inside the plane of the reflection. The dielectric film
is connected to a cantilever, which has its other end above the
surface under investigation. We consider the atomic dynamics along
$z$-axis, normal to the surface of the mirror. } \label{one}
\end{figure}

In order to realize the Recurrence Tracking microscope, we place a
cloud of cold atoms, trapped in a magneto-optic trap, above an
atomic mirror. The mirror for the atomic de Broglie waves is
obtained by the total internal reflection of a monochromatic laser
light from a dielectric film. The dielectric film is attached to the
cantilever, which has its other end above the surface under
investigation, as shown in Fig.~\ref{one}.
\begin{figure*}%\includegraphics{potential_inter.eps}
\caption{The net potential seen by an atom in the presence of an
optical field and a gravitational field (left), and its evolution
over two bounces (right): On switching off the MOT, the atom starts
its motion from an initial height $z_0$ at time $t=0$. It moves under the
influence of the linear gravitational potential, $V_{gr}=mgz$,
towards the evanescent wave atomic mirror and experiences a constant
attractive force. Close to the surface of the mirror the effect of
the evanescent light field is dominant. The atom, due to the
exponentially increasing optical potential $V_{opt}=V_0 e^{-\kappa
z}$, experiences an exponentially increasing repulsive force and bounces back.
Both the potentials together make an atomic trampoline or a
gravitational cavity for the atom. On the right, we display the
evolution of the material wave packet over two bounces in position
space as a function of time. } \label{two}
\end{figure*}

At the onset of the experiment, the magneto-optic trap is switched
off, and we let the atoms move towards the atomic mirror under the
influence of the gravitational field. We consider that the
frequency, $\nu$, of the optical field which makes the atomic mirror
is tuned far from the transition frequency, $\nu_0$, between any two
atomic levels. The probability of spontaneous
emission~\cite{kn:seif} is
%\begin{equation}
$P_{sp}= \gamma\frac{\Omega^2_{max}}{4\delta^2}\tau_{ref},$
%\end{equation}
where, $\delta=\nu-\nu_0$ is the amount of atom-field detuning,
$\gamma$ is the decay constant of the higher state, and
$\tau_{ref}=2/\kappa v_z$ is the characteristic time of the
atom-field interaction during the process of reflection off the
atomic mirror. The quantity $v_z$ describes the velocity of the atom
along the gravitational field in the $z$ direction. Hence, for a
large detuning, $\delta$, the probability of finding the atom in the
excited state becomes smaller. For the reason it becomes possible to
neglect spontaneous emission in experiments under the condition of a
large atom-field detuning~\cite{kn:sten}.

The atomic mirror is made up of an evanescent wave field, ${\bf
E}(z)$, which varies with the position, as a consequence the Rabi
frequency of the atom, $\Omega={\bf d}\cdot {\bf E}(z)/\hbar$,
becomes position dependent as well. Hence, $\Omega_{max}$ expresses
the maximum Rabi frequency seen by the atom at its turning point on
the surface of the mirror. Here, the light shift due to the external
field compensates the kinetic energy of the atom of mass $m$, that
is,
%\begin{equation}
$\frac{\hbar \Omega_{max}^2}{4\delta}=mv_{z}^2/2.$
%\end{equation}

Near the dielectric surface, the atoms observe an exponentially
increasing repulsive force, $F_{opt}=V_o\kappa e^{-\kappa z}$, as
they are detuned to the blue, that is $\nu$ is larger than $\nu_0$.
Here, $\kappa^{-1}$ defines the decay length of the atomic mirror.
Therefore, away from the mirror the repulsive optical force is
negligible. However, due to the gravitational field the atom
experiences a constant gravitational force $F_g=-mg$, and is pushed
towards the mirror. Hence, the atom undergoes a bounded motion in
the presence of the optical potential and the gravitational
potential together, as shown in Fig.~\ref{two}. The bounded atomic
dynamics in so-generated gravitational cavity or atomic trampoline
is controlled by the effective Hamiltonian,
%\begin{equation}
$H=p^{2}/2m+ mgz + V_{0}e^{-\kappa z}.$
%\end{equation}
Here, $p$ describes the center-of-mass momentum along the $z$ axis.
Moreover, $m$ indicates mass of the atom, and $g$ expresses the
constant gravitational acceleration.

%\section{Tracking of Recurrences}
\section{STATIC MODE OF OPERATION}

A material wave packet, with a finite width in energy, undergoes
constructive and destructive interferences in its evolution in time.
In quantum mechanical evolution, interference plays an important
role and manifests itself in quantum recurrences. The wave-packet
follows classical evolution for a short duration of time and
reappears after a {\it classical period}. However, after a few
classical periods it spreads all over the available space following
wave mechanics and collapses. However, due to quantum dynamics it
rebuilds itself after a certain evolution time. The phenomenon is
named as the {\it quantum revival} of the wave packet and the time
at which it reappears after a collapse is {\it quantum revival
time}. At the fractions of the quantum revival time, we find partial
appearance of the initially propagated wave packet. Therefore, these
times are called {\it fractional revival}
times~\cite{Robinett,SaifPR}.

We propagate an atomic wave packet, $|\psi\rangle$, which has a
distribution over eigen states with width $\Delta n$, centered at
mean quantum number $n_0$. We express the wave packet as a
superposition in the Hilbert space defined by the eigen states of
the net potential, that is,
%\begin{equation}
$|\psi\rangle=\sum_n c_n |\phi_n\rangle,$
%\end{equation}
where, $c_n$ describes the probability amplitude of the wave packet
in the $n$th state. Moreover, $|\phi_n\rangle$ are the eigen states
of the net potential, such that, ${\hat H}|\phi_n\rangle
=E_n|\phi_n\rangle$, where ${\hat H}$ is the time independent
Hamiltonian of the system. Since there is no explicit time
dependence the wave packet, after a propagation time $t$, appears as
\begin{equation}
|\psi(t)\rangle=e^{-iHt/\hbar}|\psi_n(t=0)\rangle=\sum_n c_n
e^{-iE_nt/\hbar} |\phi_n\rangle. \label{psit}
\end{equation}
In order to study the evolution of the wave packet in the potential
we calculate the autocorrelation function, $C(t)$, defined as
\begin{equation}
C(t)=\langle \psi(0)|\psi(t)\rangle \equiv \sum_n |c_n|^2
\exp\{-iE_nt/\hbar\}.\label{auto}
\end{equation}
Here, we have used the property of orthonormality of the eigen
states. We consider that $|c_n|^2$ corresponds to a distribution
narrowly peaked around $n_0$. Therefore, we may write
Eq.~(\ref{auto}) by using Taylor's expansion for energy $E_n$ around
$n_0$. This leads us to calculate the times, $T^{(j)}_0$, at which
the system exhibits recurrences. Here $j$ is an integer. We write
these times as,
%\begin{equation}
$T^{(j)}_0=\frac{2\pi\hbar}{\frac{1}{j!} | E_n^{(j)}|  }$,
%\end{equation}
where, $E_n^{(j)}\equiv \left.\frac{\partial^jE_n}{\partial
n^j}\right|_{n=n_0}$ describes the $j$th derivative of the energy
with respect to the principal quantum number $n$, calculated at
$n=n_0$. The first term
\begin{equation}
T^{(1)}_0=2\pi\hbar\left(\left|\frac{\partial E_n}{\partial
n}\right|_{n=n_0}\right)^{-1}=2\pi\left(\left|\frac{\partial
E_I}{\partial I}\right|_{I=I_0}\right)^{-1} \label{eq:drt2}
\end{equation}
is independent of $\hbar$ in action-angle space and corresponds to
the classical period of the wave packet. Here, $I$ ($=n\hbar$) is
the classical action, therefore, $I_0=n_0\hbar$. However, the second
term of the expansion
\begin{equation}
T^{(2)}_0=2\pi\hbar\left(\frac{1}{2!} \left|\frac{\partial^2
E_n}{\partial
n^2}\right|_{n=n_0}\right)^{-1}=2\pi\left(\frac{\hbar}{2!}
\left|\frac{\partial^2 E_n}{\partial
I^2}\right|_{I=I_0}\right)^{-1}, \label{eq:qrevt}
\end{equation}
yields the quantum mechanical revival time of the wave packet in the
potential.

In order to calculate the quantum revival time for the atom in RTM,
we approximate the net potential, made up by the optical potential
and the gravity, as a triangular well potential~\cite{chen1}. The
energy of the triangular well potential is defined as
\begin{equation}
E_n=\left(\frac{m\hbar^2g^2}{2}\right)^{1/3}z_n.
\end{equation}
Here, $z_n$'s are negative zeros of Airy function, and can be
defined~\cite{kn:abra} as
%\begin{equation}
$z_n=f\left(\frac{3\pi}{2}(n-1/4)\right)$,
%\end{equation}
where
%\begin{equation}
$f(\zeta)=\zeta^{2/3}\left(1+ \frac{5}{48\zeta^2} -
\frac{5}{36\zeta^4} +\cdots\right).$
%\end{equation}
In case of large $n$, we may reduce $z_n$ to $z_n\cong
(3n\pi/2)^{2/3}$ which provides the eigen energies of the system, as
%\begin{equation}
$E_n\cong \frac{m^{1/3}}{2}\left( 3 n\pi\hbar g \right)^{2/3}.$
%\end{equation}
With the knowledge of $E_n$ at hand the quantum revival time,
$T^{(2)}_0$, is obtained with the help of Eq.~(\ref{eq:qrevt}), such
that
\begin{equation}
T^{(2)}_0=\frac{16 E_{n_0}^2}{m\pi\hbar g^2}, \label{eq:qrevt1}
\end{equation}
where, $E_{n_0}$ is the initial mean energy of the material wave
packet.

Hence, in order to probe a surface which has arbitrary structures,
we use the quantum recurrence tracking microscope in static mode. We
let the atom fall on the static atomic mirror without moving the
surface under investigation. In its evolution over the atomic mirror
for a certain fixed position of the cantilever the atom displays
quantum revival at quantum revival time, $T^{(2)}_0$, as given in
Eq.~(\ref{eq:qrevt1}).

As we slightly move the surface under study, the position of the
cantilever changes following the surface structures. This changes
the initial distance between the atomic mirror and the bouncing
atom. This leads to a different initial energy $E_{n_0}$ for the
atom, and thus a different revival time, $T^{(2)}_0$. For each new
experimentally calculated $T^{(2)}_0$, we calculate the
corresponding $E_{n_0}$, which leads to the knowledge of the
structures on the surface being probed.

\section{DYNAMICAL MODE OF OPERATION}

In case the surface under study has periodic nano-structures, we
find periodic spatial modulation of the atomic mirror as we move the
surface horizontally below the cantilever. The lower tip of the
cantilever follows the surface structures and introduces spatial
modulation to the atomic mirror at its other end. Hence, the
dynamics of an atom is controlled by an explicitly time dependent
Hamiltonian,
%\[
$H=\frac{P^{2}}{2m}+mgz+V_{0}e^{-k(z-a\sin\omega t)}$,
%\]
where, $a$ describes the amplitude of the spatial modulation which
corresponds to the height of the periodic structures and $\omega$
defines the frequency at which the structures appear.

In the presence of the spatial modulation of the atomic mirror, the
bouncing atom displays collapse and revival after a definite period
of time. We~\cite{saif} calculate the time of the quantum revival as
\begin{equation}
T^{(2)}_{\lambda }=T^{(2)}_0\left[ 1-\frac{1}{8}\left(
\frac{mga}{E_{n_0}} \right)
^{2}\frac{3(1-r)^{2}+\tilde{a}_{{}}^{2}}{[\left( 1-r\right)
^{2}-\tilde{a}^{2}]^{3}}\right] ,\label{t2}
\end{equation}
where, $r\equiv (E_{N}/E_{n_0})^{1/2}$ and $\tilde{a}=r^{2}\hbar
\omega /4 E_{n_0}$ are dimensionless parameters. The time
$T^{(2)}_0$, expressed in Eq.~(\ref{eq:qrevt1}), defines the quantum
revival time for the atomic wave packet in the {\it absence} of the
modulation of the atomic mirror, that is, on the static surface. The
time of the quantum revival in the {\it presence} of modulation, as
given in Eq.~(\ref{t2}), depends upon the frequency, $\omega$, and
the height of the periodic structures, $a$.

We measure $a$ by considering that the material wave packet is
released from the MOT, with a large initial mean energy $E_{n_0}$,
such that $(1-r)^{2}>\tilde{a}^{2}$. Hence, the time of the quantum
revival $T^{(2)}_{\lambda}$ becomes,
\[
T^{(2)}_{\lambda }=T^{(2)}_0\left[ 1-\frac{3}{8}\left( \frac{mga}{E_{n_0}}%
\right) ^{2}\frac{1}{\left( 1-r\right) ^{4}}\right] ,
\]
which provides the value of $a$, expressing the height of the
periodic structures as,
\begin{equation}
a=\sqrt{\frac{8}{3}}\frac{E_{n_0}}{mg}\left( 1-r\right) ^{2}\left[
1-\frac{T^{(2)}_{\lambda }}{T^{(2)}_0}\right] ^{1/2}.\label{height}
\end{equation}
\begin{figure}\includegraphics[scale=0.7]{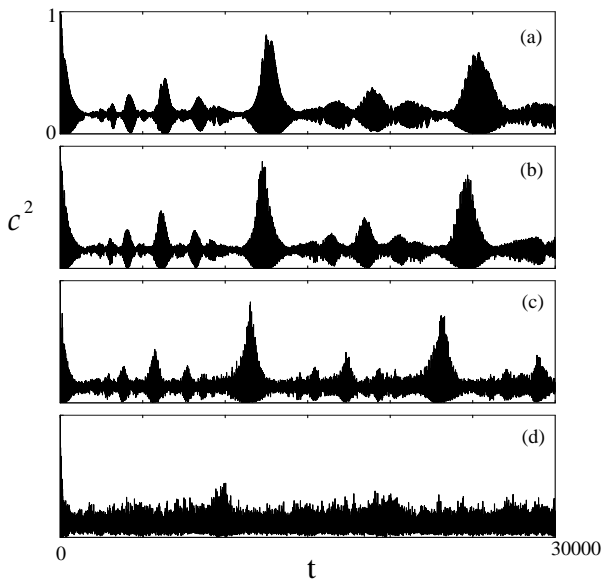}
\caption{We propagate a cloud of Cs atoms centered at mean quantum
number $n_0=176.16$, with a width $\Delta z=0.28\mu m$, from a
height $z_0=20.1\mu m$~\cite{saif}. We calculate the time
of quantum revivals numerically for different values of initial
modulations. We substitute these values in Eqs.~(\ref{height}) and
(\ref{alphat}) and get the corresponding values of $a$ as (a) 0, (b)
11.4$nm$ (c) 23$nm$ (d) 46$nm$ and frequency as $\omega=2\pi$x$
1KHz$~\cite{kn:sten}. If the surface under study is moved below
cantilever with a velocity of $1\mu m/sec$ the spacing between the
surface structures is noted as $1nm$. } \label{three}
\end{figure}

Knowledge of the value of $a$ helps us to find the frequency,
$\omega$, of the appearance of the periodic structures. We may
express Eq.~(\ref{t2}) as,
\begin{equation}
\left[ \left( 1-r\right) ^{2}-\tilde{a}^{2}\right] ^{3}\alpha
_{T}-\left[ 3\left( 1-r\right) ^{2}+\tilde{a}^{2}\right]
=0,\label{roots}
\end{equation}
where,
\begin{equation}
\alpha _{T}=8\left( \frac{E_{n_0}}{mga}\right) ^{2}\left[ 1-%
\frac{T^{(2)}_{\lambda }}{T^{(2)}_0}\right] .\label{alphat}
\end{equation}
We find the values of $T^{(2)}_0$ and $T^{(2)}_{\lambda}$
experimentally, and substitute them in Eq.~(\ref{alphat}) to obtain
the value of $\alpha_T$. Thus, we find $\tilde{a}$ as the roots of
Eq.~(\ref{roots}), which yields the frequency $\omega$, as
%\begin{equation}
$\omega =4E_{n_0}\tilde{a}/r^2\hbar $.
%\label{frequency}
%\end{equation}
Hence, the measurement of the quantum revival times, $T^{(2)}_0$ and
$T^{(2)}_{\lambda}$, and the height $a$, help to measure the
frequency, $\omega$. This immediately leads us to calculate the
spacing between the surface structures as we know the velocity at
which the surface is moved horizontally below the cantilever.

We calculate the square of the auto correlation function and display our numerical results in Fig.~\ref{three}. We consider a
cloud of cesium atoms initially in a Gaussian distribution centered
at mean quantum number $n_0=176.16$, with a width $\Delta z=0.28\mu
m$, which is propagated from an initial height $z_0=20.1\mu
m$~\cite{Ovchinnikov1997}. We calculate the time of quantum revivals
numerically for different modulations of the atomic mirror. On
substituting the values in Eqs.~(\ref{height}) and (\ref{alphat}) we
obtain the height $a$, and the frequency $\omega$ at which the
periodic structures appear on the surface.
%Since the surface under study was moved
%below cantilever with a velocity of $1\mu m$ we find the spacing
%between the structures as $1nm$.
%A very good agreement is found between the revival time obtained
%from our analytical calculations and the numerical results, and
%displayed in Fig.~\ref{three}.

%The suggested QSM is having advantages over existing techniques of STM and
%AFM. We can use QSM to study any structure on a surface including periodic
%and non-periodic, moreover, the QSM is applicable to the study of conducting
%surfaces and non-conducting surfaces as well.

We suggest that the Recurrence tracking microscope opens new
horizons to study and incorporate effects of atomic coherence and
interference. As discussed above, the RTM has clear advantages
over STM and AFM. The bouncing atoms reflect back {\it above} the
dielectric surface and do not influence the dielectric surface or
cantilever, this may increase the stability of the system. The
ability to scan all kind of surfaces is another credit to the
device. In addition, RTM in dynamical mode scans the periodic
structures more swiftly~\cite{saifpatent}.

\section{ACKNOWLEDGMENT}

The author submits his thanks to Prof. Pierre Meystre for his hospitality at the Department of Physics, the University of Arizona.

%\end{multicols}

\end{document}